\documentclass[final]{elsart}
\usepackage{graphicx}
\usepackage{amsmath}
\usepackage{amssymb}
\usepackage{mycommands}
\usepackage{subfigure}
\usepackage{epsfig}
\usepackage{lscape}

\journal{Physics Letters B}

\def\nm {\mbox{\boldmath $\nu_\mu$}} 
\def\anm{\mbox{\boldmath $\bar\nu_\mu$}} 
\def\ne {\mbox{\boldmath $\nu_e$}} 
\def\ane{\mbox{\boldmath $\bar\nu_e$}}

\def\cohp{\boldmath {Coh$\pi^0$} }
\def\piz{\boldmath {$\pi^0 $} }

\def\mgg{\boldmath {M$_{\gamma \gamma}$} }
\def\gam{\boldmath {$\gamma $} }
\def\gf{\boldmath {$\gamma 1$} }
\def\gs{\boldmath {$\gamma 2$} }
\def\A{\boldmath {${\cal A}$} }
\def\zt{\boldmath {$\zeta$}}

\def\ztf{\boldmath {$\zeta_{\gamma 1}$}}
\def\zts{\boldmath {$\zeta_{\gamma 2}$}}
\def\thfs{\boldmath {$\Theta_{1 2}$}}

\begin{document}

\begin{frontmatter}

\title{
A Measurement of Coherent  Neutral Pion Production in 
Neutrino Neutral Current Interactions in the NOMAD
Experiment}

\author[19]{C.T.~Kullenberg}
\author[19]{S.R.~Mishra}
\author[19]{M.B.~Seaton}
\author[19]{J.J.~Kim}
\author[19]{X.C.~Tian}
\author[19]{A.M.~Scott}
\author[12]{M.~Kirsanov}
\author[19]{R.~Petti}
\author[25]{S.~Alekhin}
\author[14]{P.~Astier}
\author[8]{D.~Autiero}
\author[18]{A.~Baldisseri}
\author[13]{M.~Baldo-Ceolin}
\author[14]{M.~Banner}
\author[1]{G.~Bassompierre}
\author[9]{K.~Benslama}
\author[18]{N.~Besson}
\author[8,9]{I.~Bird}
\author[2]{B.~Blumenfeld}
\author[13]{F.~Bobisut}
\author[18]{J.~Bouchez}
\author[20]{S.~Boyd\thanksref{Now1}}
\thanks[Now1]{Now at University of Warwick, UK}
\author[3,24]{A.~Bueno}
\author[6]{S.~Bunyatov}
\author[8]{L.~Camilleri}
\author[10]{A.~Cardini}
\author[15]{P.W.~Cattaneo}
\author[16]{V.~Cavasinni}
\author[8,22]{A.~Cervera-Villanueva}
\author[11]{R.~Challis}
\author[6]{A.~Chukanov}
\author[13]{G.~Collazuol}
\author[8,21]{G.~Conforto \thanksref{Deceased}}
\thanks[Deceased]{Deceased}
\author[15]{C.~Conta}
\author[13]{M.~Contalbrigo}
\author[10]{R.~Cousins}
\author[9]{H.~Degaudenzi}
\author[8,16]{A.~De~Santo}
\author[16]{T.~Del~Prete}
\author[8]{L.~Di~Lella \thanksref{Now2}}
\thanks[Now2]{Now at Scuola Normale Superiore, Pisa, Italy}
\author[8]{E.~do~Couto~e~Silva}
\author[14]{J.~Dumarchez}
\author[20]{M.~Ellis\thanksref{Now3}}
\thanks[Now3]{Now at Brunel University, Australia}
\author[3]{G.J.~Feldman}
\author[15]{R.~Ferrari}
\author[8]{D.~Ferr\`ere}
\author[16]{V.~Flaminio}
\author[15]{M.~Fraternali}
\author[1]{J.-M.~Gaillard}
\author[8,14]{E.~Gangler}
\author[5,8]{A.~Geiser}
\author[5]{D.~Geppert}
\author[13]{D.~Gibin}
\author[8,12]{S.~Gninenko}
\author[19]{A.~Godley}
\author[8,22]{J.-J.~Gomez-Cadenas}
\author[18]{J.~Gosset}
\author[5]{C.~G\"o\ss ling}
\author[1]{M.~Gouan\`ere}
\author[8]{A.~Grant}
\author[7]{G.~Graziani}
\author[13]{A.~Guglielmi}
\author[18]{C.~Hagner}
\author[22]{J.~Hernando}
\author[3]{P.~Hurst}
\author[11]{N.~Hyett}
\author[7]{E.~Iacopini}
\author[9]{C.~Joseph}
\author[9]{F.~Juget}
\author[11]{N.~Kent}
\author[6]{O.~Klimov}
\author[8]{J.~Kokkonen}
\author[12,15]{A.~Kovzelev}
\author[1,6]{A. Krasnoperov}
\author[12]{S.~Kulagin}
\author[13]{S.~Lacaprara}
\author[14]{C.~Lachaud}
\author[23]{B.~Laki\'{c}}
\author[15]{A.~Lanza}
\author[4]{L.~La Rotonda}
\author[13]{M.~Laveder}
\author[14]{A.~Letessier-Selvon}
\author[14]{J.-M.~Levy}
\author[19]{J.~Ling} 
\author[8]{L.~Linssen}
\author[23]{A.~Ljubi\v{c}i\'{c}}
\author[2]{J.~Long}
\author[7]{A.~Lupi}
\author[6]{V.~Lyubushkin}
\author[7]{A.~Marchionni}
\author[21]{F.~Martelli}\
\author[18]{X.~M\'echain}
\author[1]{J.-P.~Mendiburu}
\author[18]{J.-P.~Meyer}
\author[13]{M.~Mezzetto}
\author[11]{G.F.~Moorhead}
\author[6]{D.~Naumov}
\author[1]{P.~N\'ed\'elec}
\author[6]{Yu.~Nefedov}
\author[9]{C.~Nguyen-Mau}
\author[17]{D.~Orestano}
\author[17]{F.~Pastore}
\author[20]{L.S.~Peak}
\author[21]{E.~Pennacchio}
\author[1]{H.~Pessard}
\author[8]{A.~Placci}
\author[15]{G.~Polesello}
\author[5]{D.~Pollmann}
\author[12]{A.~Polyarush}
\author[11]{C.~Poulsen}
\author[6,14]{B.~Popov}
\author[13]{L.~Rebuffi}
\author[24]{J.~Rico}
\author[5]{P.~Riemann}
\author[8,16]{C.~Roda}
\author[8,24]{A.~Rubbia}
\author[15]{F.~Salvatore}
\author[6]{O.~Samoylov}
\author[14]{K.~Schahmaneche}
\author[5,8]{B.~Schmidt}
\author[5]{T.~Schmidt}
\author[13]{A.~Sconza}
\author[11]{M.~Sevior}
\author[1]{D.~Sillou}
\author[8,20]{F.J.P.~Soler}
\author[9]{G.~Sozzi}
\author[2,9]{D.~Steele}
\author[8]{U.~Stiegler}
\author[23]{M.~Stip\v{c}evi\'{c}}
\author[18]{Th.~Stolarczyk}
\author[9]{M.~Tareb-Reyes}
\author[11]{G.N.~Taylor}
\author[6]{V.~Tereshchenko}
\author[12]{A.~Toropin}
\author[14]{A.-M.~Touchard}
\author[8,11]{S.N.~Tovey}
\author[9]{M.-T.~Tran}
\author[8]{E.~Tsesmelis}
\author[20]{J.~Ulrichs}
\author[9]{L.~Vacavant}
\author[4]{M.~Valdata-Nappi\thanksref{Now4}}
\thanks[Now4]{Now at Univ. of Perugia and INFN, Perugia, Italy}
\author[6,10]{V.~Valuev}
\author[14]{F.~Vannucci}
\author[20]{K.E.~Varvell}
\author[21]{M.~Veltri}
\author[15]{V.~Vercesi}
\author[8]{G.~Vidal-Sitjes}
\author[9]{J.-M.~Vieira}
\author[10]{T.~Vinogradova}
\author[3,8]{F.V.~Weber}
\author[5]{T.~Weisse}
\author[8]{F.F.~Wilson}
\author[11]{L.J.~Winton}
\author[19]{Q.~Wu\thanksref{Now5}}
\thanks[Now5]{Now at Illinois Institute of Technology, USA}
\author[20]{B.D.~Yabsley}
\author[18]{H.~Zaccone}
\author[5]{K.~Zuber}
\author[13]{P.~Zuccon}

\address[1]{LAPP, Annecy, France}
\address[2]{Johns Hopkins Univ., Baltimore, MD, USA}
\address[3]{Harvard Univ., Cambridge, MA, USA}
\address[4]{Univ. of Calabria and INFN, Cosenza, Italy}
\address[5]{Dortmund Univ., Dortmund, Germany}
\address[6]{JINR, Dubna, Russia}
\address[7]{Univ. of Florence and INFN,  Florence, Italy}
\address[8]{CERN, Geneva, Switzerland}
\address[9]{University of Lausanne, Lausanne, Switzerland}
\address[10]{UCLA, Los Angeles, CA, USA}
\address[11]{University of Melbourne, Melbourne, Australia}
\address[12]{Inst. for Nuclear Research, INR Moscow, Russia}
\address[13]{Univ. of Padova and INFN, Padova, Italy}
\address[14]{LPNHE, Univ. of Paris VI and VII, Paris, France}
\address[15]{Univ. of Pavia and INFN, Pavia, Italy}
\address[16]{Univ. of Pisa and INFN, Pisa, Italy}
\address[17]{Roma Tre University and INFN, Rome, Italy}
\address[18]{DAPNIA, CEA Saclay, France}
\address[19]{Univ. of South Carolina, Columbia, SC, USA}
\address[20]{Univ. of Sydney, Sydney, Australia}
\address[21]{Univ. of Urbino, Urbino, and INFN Florence, Italy}
\address[22]{IFIC, Valencia, Spain}
\address[23]{Rudjer Bo\v{s}kovi\'{c} Institute, Zagreb, Croatia}
\address[24]{ETH Z\"urich, Z\"urich, Switzerland}
\address[25]{Inst. for High Energy Physics, 142281, Protvino, Moscow, Russia}

\begin{abstract}
We present a study of exclusive neutral pion production in 
neutrino-nucleus Neutral Current interactions 
using data from the NOMAD experiment
at the CERN SPS. The data correspond  to $1.44 \times 10^6$ 
muon-neutrino Charged Current
interactions in the energy range 
$2.5 \leq E_{\nu} \leq 300$~GeV. Neutrino events with 
only one visible  $\pi^0$ in the final state 
are expected to result from two 
Neutral Current processes: 
coherent $\pi^0$ production, 
{\boldmath $\nu + {\cal A} \rightarrow \nu + {\cal A} + \pi^0$} 
and single $\pi^0$ production in neutrino-nucleon scattering. 
The signature of coherent $\pi^0$ production 
is an emergent $\pi^0$ almost collinear with the
incident neutrino while $\pi^0$'s 
produced in neutrino-nucleon deep inelastic
scattering have larger transverse momenta. 
In this analysis all relevant
backgrounds to the coherent $\pi^0$ production 
signal are measured using data themselves. 
Having determined the backgrounds, 
and using the Rein-Sehgal model for the 
coherent $\pi^0$ production 
to compute the detection efficiency, 
we obtain 
{\boldmath $4630 \pm 522 (stat) \pm 426 (syst)$} 
corrected coherent-$\pi^0$ events 
with  $E_{\pi^0} \geq 0.5$~GeV. We measure 
{\boldmath 
$\sigma (\nu {\cal A} \rightarrow \nu {\cal A} \pi^0) = 
\left [ 72.6 \pm 8.1(stat) \pm 6.9(syst)  \right ] \times 10^{-40} 
cm^2/nucleus$}. 
This is the most precise 
measurement of the coherent $\pi^0$ production to date.

\end{abstract}


\end{frontmatter}

\section{Motivation}
\label{sec-intro}
\begin{figure}
\begin{center}
\includegraphics[width=0.5\textwidth]
{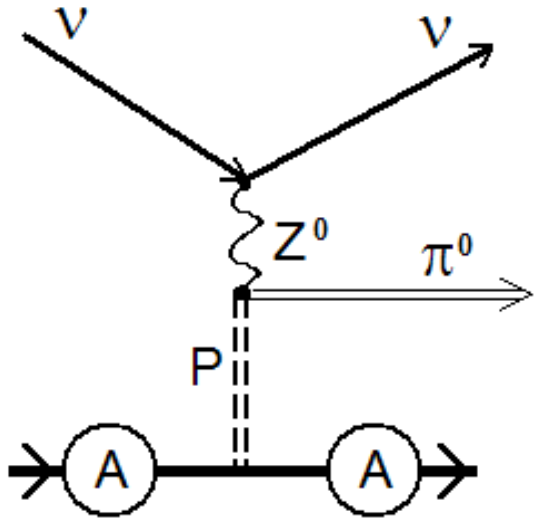}
\caption
{Diagram of the \cohp\ process, 
{\boldmath $\nu + {\cal A} \rightarrow \nu + {\cal A} + \pi^0$}. }
\label{fig-feynman}
\end{center}
\end{figure}

Precise measurement of $\pi^0$ production 
when a neutrino scatters coherently off a 
target nucleus, 
{\boldmath $\nu + {\cal A} \rightarrow \nu + {\cal A} + \pi^0$}, 
depicted in Figure~\ref{fig-feynman}, 
is challenging: the cross-section ($\sigma$)
of coherent-$\pi^0$ (\cohp) is 0.003  
of the inclusive neutrino charged current (CC) interactions 
at $E_\nu \simeq 25$~GeV~\cite{Rein:1982pf}; 
the single $\pi^0$ is notoriously refractory to accurate 
identification in neutrino detectors. Consequently 
the past cross-section measurements of \cohp\ 
have been poor, 
with a precision no better than $\simeq 30\%$
~\cite{EXAP,EXGGM,EXCHARM,EXSKAT,EX15FT}; 
recently the MiniBOONE experiment
has reported the fraction of \cohp\ in all exclusive NC $\pi^0$ 
production~\cite{EXMB} . 
This challenge is the primary motivation 
for the present analysis. The second motivation is utilitarian. 
Since \cohp\ is almost collinear 
with the incident neutrino, in massive neutrino detectors 
a \cohp\ event will manifest itself as a 
forward electromagnetic shower 
posing a background for the \ne-induced signal.
This is relevant to the long baseline experiments 
searching for \ne\ appearance with the 
purpose of measuring the mixing angle $\Theta_{13}$.
A precise measurement of \cohp, 
although conducted at energies higher than those of the 
long baseline projects at Fermilab (MINOS/NO$\nu$A), 
will constrain the error on a model-prediction of this  
background to the \ne\ appearance. 
Finally, the study of coherent pion production provides an 
insight into the structure of the weak hadronic 
current~\cite{Rein:1982pf, Belkov:1986hn}, and offers  
a test of the partially conserved axial-vector current hypothesis 
(PCAC)~\cite{ADLER}. Ref.~\cite{Kopeliovich:1992ym} 
presents  an excellent review of these topics. 

A coherent interaction, 
Figure~\ref{fig-feynman}, 
where no charge or isospin is 
exchanged between the $\nu$ and 
the target nucleus (${\cal A}$) which 
recoils without breakup, leads 
to an enhancement in the cross-section. In the \cohp\ process 
the interaction is mediated by a pomeron-like particle 
bearing the quantum number of the vacuum.  The 
cross section is dominated by the axial vector current. 
The contribution of the isovector current to the \cohp\ process 
is minimal where $Z^0$ can be viewed as a $\rho$ meson which 
produces a $\pi^0$ exchanging an isoscalar $\omega$ with 
${\cal A}$. This minimal contribution of the isovector 
current to the \cohp\ arises from two reasons: 
(a) the cross section of  the isovector $\rho$-${\cal A}$ 
interaction is zero in the forward direction, a direction preferred by 
the nuclear form factor; and (b) the vector component has 
a contribution proportional to $(1-2\sin^2\theta_W)^2$ 
reducing the isovector contribution further, 
the net reduction with respect to the axial part being a factor of 
3.5. The PCAC hypothesis stipulates 
that for zero-momentum transfer ($Q^2=0$, where $Q^2$ 
is the negative of the square of the four-momentum 
transfer from the  incident neutrino to the target), 
 the $\nu$-${\cal A}$ cross section can be 
related to the $\pi$-${\cal A}$ cross section.
The $\nu$-${\cal A}$ 
cross section in the forward direction is related to the strong 
$\pi$-${\cal A}$  interaction as follows: 

\begin{equation} 
\left [ \frac{d^3\sigma (\nu {\cal A} \rightarrow \nu {\cal A} \pi^0)}{dxdydt} \right ]_{Q^2=0}=
\frac{G^2ME_\nu}{\pi^2} \frac{1}{2} f^2_\pi (1-y) 
\left [ \frac {d\sigma(\pi {\cal A} \rightarrow \pi {\cal A})}{dt} \right ]_{yE_\nu=E_\pi} 
\label{eq-rscohq0}
\end{equation}

\noindent
In Equation~(\ref{eq-rscohq0}) 
$G$ is the Fermi coupling constant, 
$M$ is the nucleon mass, $x=Q^2/2M\nu$ and 
$y=\nu/E_\nu$, where $\nu$ is the energy 
of the hadronic system in the final state, 
are the standard scaling variable, 
and $f_\pi=0.93\,m_\pi$ is the pion decay constant. 
The variable $t$ quantifies the coherence (forwardness) and is 
defined as $t=p^2_T=(q-P_\pi)^2$, i.e. the square of 
the four-momentum transfer to the nucleus. 
In a neutral current (NC) event since the emergent 
neutrino remains invisible,  $|t|$ cannot be measured. 
Instead the very small transverse momentum expected 
in a coherent interaction can be quantified using the variable
$\zeta$ defined as: 
$\zeta_{\pi^0}=E_{\pi^0} \left [ 1-\cos(\theta_{\pi^0}) \right ].$ 
This variable has the property that its
distribution depends weakly on the incident neutrino energy. 

For low but non-zero $Q^2$ values, the hadron 
dominance model~\cite{HDM} provides a guide 
to extend the cross section 
formula for the \cohp-like process. The $Z^0$ boson can be 
viewed as a superposition of axial vector and vector currents. 
These compose the weak hadronic current. 

\section{Beam and Detector}
\label{sec-nomad}

The Neutrino Oscillation MAgnetic 
Detector (NOMAD) experiment at CERN used a 
neutrino beam  ~\cite{CERN-BEAM}
produced by 
the 450~GeV protons from the 
Super Proton Synchrotron (SPS) incident on
 a beryllium target and producing  
secondary $\pi^{\pm}$, $K^{\pm}$, and $K^0_L$ mesons. 
The positively charged mesons were focussed by 
two magnetic horns into a 
290~m long evacuated decay pipe. Decays of  
$\pi^{\pm}$,  $K^{\pm}$, and $K^0_L$  
produced the SPS neutrino beam. 
The average neutrino flight path to  
NOMAD was 628~m, the detector being 
836~m downstream of the Be-target.  
The SPS beamline  and the neutrino flux incident 
at NOMAD are described in~\cite{NOMAD-FLUX}.  
The $\nu$-flux in NOMAD is constrained by the 
$\pi^{\pm}$ and $K^{\pm}$ production measurements in 
proton-Be collision by the SPY experiment 
~\cite{SPY1, SPY2, SPY3} and by an 
earlier measurement conducted by 
Atherton {\it et al.}~\cite{ATHERTON}. 
The  $E_\nu$-integrated relative composition of 
\nm:\nmb:\nel:\neb\ CC events, 
constrained $in$ $situ$ by the 
measurement of CC-interactions 
of each of the  neutrino species,  is 
$1.00: 0.025: 0.015:0.0015$. Thus,  95\% of $\nu$-events, 
are due to \nm-interactions with a small  \nmb-contamination.

\begin{landscape}
\begin{figure}
\begin{center}
\includegraphics[scale=1.00]
{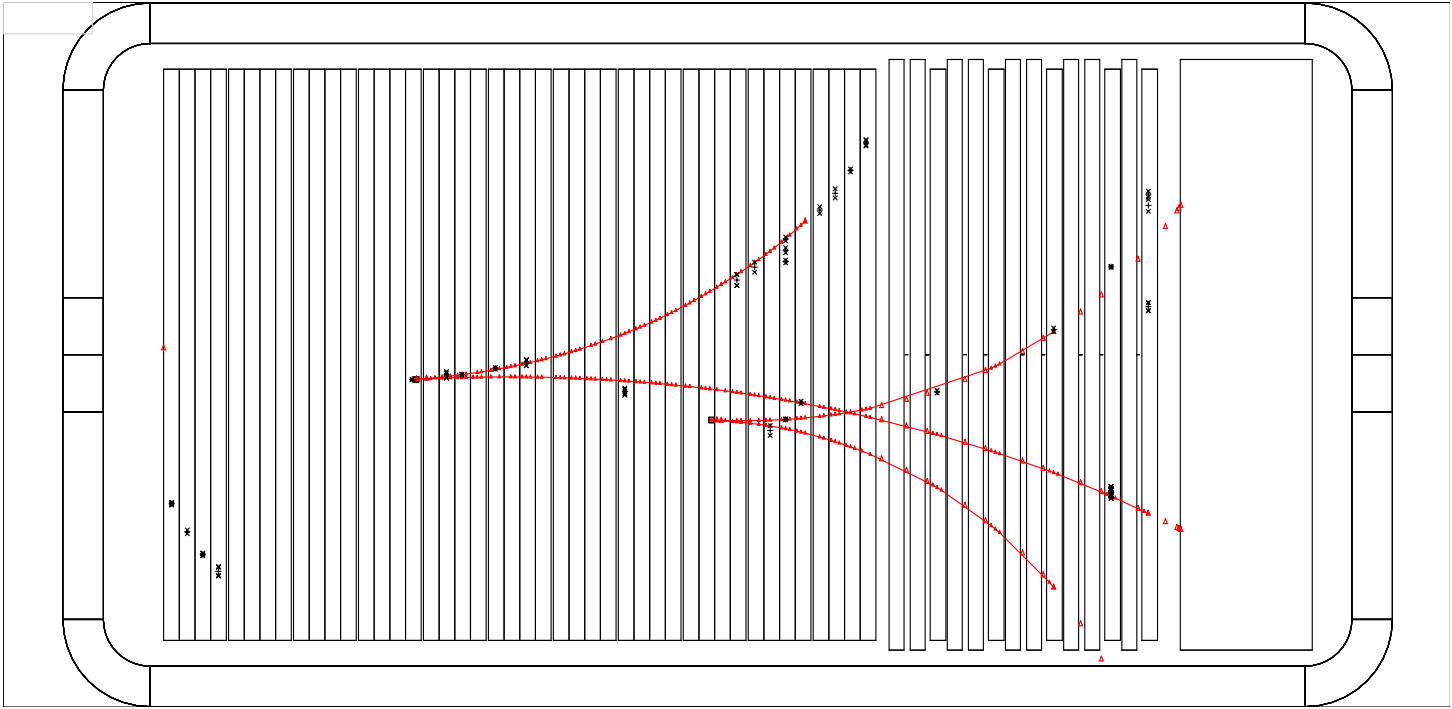}
\caption{Schematic of the DC tracker and a 
coherent $\pi^0$ event candidate in NOMAD where both 
photons from the $\pi^0$ decay convert in the DC's. 
The red crosses represent drift chamber digitizations that 
are used in the track-reconstruction, 
whereas the black ones are not. 
The upstream (\gf)  and downstream (\gs) momentum 
vectors when extrapolated upstream 
intersect within the fiducial volume.
}
\label{fig-evtpi01}
\end{center}
\end{figure}
\end{landscape}

The NOMAD experiment was designed to search for 
\mutotau\ oscillations at $\Delta m^2 \geq 5$~eV$^2$, 
and  in large $\Delta m^2$ range it set 
stringent limit~\cite{NOMAD-NMNT} on this search, 
along with the CHORUS experiment~\cite{CHORUS-NMNT}.  
The NOMAD apparatus~\cite{NOMAD-NIM} 
was composed of several sub-detectors. The active 
target comprised 132 planes of $3 \times 3$~m$^2$ drift chambers (DC)    
with an average density similar to that of  liquid 
hydrogen (0.1~gm/cm$^3$). 
On average, the equivalent material in the DC 
encountered by 
particles produced in a $\nu$-interaction  
was about half a radiation  length 
and a quarter of an hadronic interaction length ($\lambda$).  
The fiducial mass of the NOMAD DC-target,  2.7 tons, was  
composed  primarily of carbon (64\%), oxygen (22\%), nitrogen (6\%), 
and hydrogen (5\%) yielding an effective atomic number, 
\A=12.8, similar to carbon.
Downstream of the DC, there were nine modules of transition radiation 
detectors (TRD), followed by a preshower (PRS) and a lead-glass 
electromagnetic calorimeter (ECAL). 
The ensemble of DC, TRD, and PRS/ECAL was placed within 
a dipole magnet providing a 0.4~T magnetic field orthogonal 
to the neutrino beam line. 
Two planes of scintillation counters, $T_1$ and $T_2$, 
positioned upstream and downstream of the TRD, 
provided the trigger in combination with an 
anti-coincidence signal, ${\overline V}$, 
from the veto counter upstream and outside the magnet. 
Downstream  of the magnet was a hadron calorimeter, 
followed by two muon-stations each comprising large area 
drift chambers and separated by an iron filter 
placed at 8- and 13-$\lambda$'s downstream of 
the ECAL, that provided a clean identification of the muons. 
The schematic of the detector 
in the Y-Z view is shown in Figure~\ref{fig-evtpi01}.
The charged tracks in the DC were measured with an 
approximate  momentum ($p$)  resolution of  
$\sigma_p/p = 0.05/\sqrt{L} \oplus 0.008p/\sqrt{L^5}$  
($p$ in GeV/$c$ and $L$ in meters) 
with unambiguous charge separation in the energy range of interest. 
The detailed individual reconstruction 
of each charged and neutral track and their  
precise momentum vector measurement  
enabled a quantitative description of 
the event kinematics: the strength and 
basis of NOMAD analyses. 
The experiment recorded over 1.7 million 
neutrino interactions in its active drift-chamber  (DC) target. 
These data are unique in that they constitute the largest
high resolution neutrino data sample with 
accurate identifications of \nm, \nmb, \nel, and \neb\  charged 
current interactions  in the energy range 
${\cal O}(1) \leq E_\nu \leq 300$~GeV.  
In addition, the experiment recorded over 2 million  
$\nu$-interactions in the Al-coil  
and over 20 million in the Fe-scintillator calorimeter, 
both upstream of the active-DC target.

\newpage

\section{The \cohp\ Signature and Models}
\label{sec-sig-rsbk}

The signature for \cohp\  is a single forward $\pi^0$ 
and nothing else. The \piz\ will promptly decay into 
two forward photons ($\gamma$). In massive neutrino 
detectors the signal will manifest itself as an electromagnetic 
shower, short and compact, with a forward direction. 
The accompanying irreducible backgrounds will be 
\ne, \ane, and $\nu$-NC events dominated by 
$\pi^0$'s. In NOMAD, however, the \cohp\ 
signal will reveal two distinct photons. 
The photons will either both convert in the DC target,  
or one of the photons will convert in the tracker and the other 
will be measured in the electromagnetic calorimeter (ECAL), 
or both photons will be measured in the ECAL.  
In this analysis we focus on the event 
sample where both photons convert in the DC target.  
Figure~\ref{fig-evtpi01} shows such an event. The momenta 
of the associated $e^-$ and $e^+$ are measured  in the magnetic field. 
Each event thus provides a complete $\pi^0$-momentum vector. 
We use the Rein-Sehgal (RS) model~\cite{Rein:1982pf}  
to simulate the \cohp\ interaction in the NOMAD detector.
As a check we also simulated the \cohp\ interaction following the 
Belkov-Kopeliovich (BK)~\cite{Belkov:1986hn} model. 
The \piz\ reconstruction efficiency computed using 
the BK model is similar to that determined by the RS model. 

Recently a set of new \cohp\ calculations has  
been proposed (see~\cite{Singh:2006bm}, 
\cite{AlvarezRuso:2007it}, and~\cite{Paschos:2005km}). 
They focus on \cohp\ production 
in low-energy neutrino interaction (${\cal {O}} (1)$~GeV). 
However, the present \cohp\ measurement at 
an average $E_\nu \simeq 25$~GeV,  more 
precise by about a factor of three than currently available,     
could be used to constrain parameters used in 
these calculations.

\section{Selection of  Exclusive $2$-$\gamma$ Events}
\label{sec-sel}
We select events with two converted 
photons in the DC target. The analysis 
uses the entire NOMAD data and the 
associated Monte Carlo (MC) samples 
as described in ~\cite{NOMAD-XSEC}. 
The number of fully corrected \nm-CC 
in the standard fiducial volume of NOMAD is 
$1.44 \times 10^6$ events: the denominator for 
the present measurement. 
The NC-DIS sample,  defined by requiring that 
the generated invariant hadronic mass squared ($W^2$) 
be $\geq 1.96$~GeV$^2$,  
is normalized to $0.53 \times 10^6$ 
events which corresponds to 0.37 of the \nm-CC. The 
NC-Resonance ($W^2 \leq 1.96$) 
sample is set at 3.5\% of the NC-DIS. 
The MC sample  specific to this analysis  is the 
RS \cohp\ simulation. Motivated by the \nm-induced 
coherent-$\pi^+$ cross sections 
presented in~\cite{Belkov:1986hn} and the fact 
that the NC/CC coherent pion cross section ratio should be (1/2), 
the \cohp\ sample is  normalized to 5000 events  
with generated $E_{\pi^0} \geq 0.5$~GeV. 
The large sample of data and those of the NC and CC 
deep inelastic scattering (DIS) MC events are 
subjected to a preselection. The preselection includes 
the following requirements: (a) the presence of one 
converted photon whose reconstructed 
conversion point is defined as the event vertex ($X$, $Y$, $Z$); 
(b) no identified muons; (c) vertex coordinates 
of the converted photon within the fiducial volume, 
$|X,(Y-5)|\leq 130$~cm and $Z_{Min} \leq Z \leq 405$~cm where 
$Z_{Min}$ depends upon the detector configuration (see 
~\cite{NOMAD-XSEC} for detail); 
(d) the invariant mass ($M_{ee}$) of 
the $e^-$ and $e^+$   less than 100~MeV/$c^2$ which  
selects both the  converted photons --- the upstream being \gf,  
and the downstream being \gs ---,  
with  95\% purity and  97\% efficiency.  
The preselection reduces the data and the NC-MC samples by 
a factor of about a hundred. 

The cuts for the final selection of the \cohp\ events are set 
to maximize the selection efficiency 
of two photon conversions in the DC tracker.  
The cuts are optimized to reduce the 
NC-DIS background while keeping the \cohp\ signal high. 
We also look at about 10\% of the data 
to check the efficacy of cuts used 
in reducing the  background induced 
by $\nu$-interactions occurring outside the fiducial 
volume --- the outside background (OBG).  
The remaining 
data have no 
influence on the choice of the cuts. 
The results presented here include the entire data sample. 
Among the generated \cohp,  
only about 29\% of events trigger the apparatus. The 
loss arises from the non-converted photons ($\simeq 50\%$)  
and, among the converted photons,  from 
the $e^-/e^+$ tracks that do not reach 
the downstream trigger counters ($\simeq 20\%$). 

The final event selection follows the preselection 
cuts with more stringent requirement. The $M_{ee}$ 
cut is tightened to $50$~MeV/$c^2$ which increases 
the photon conversion 
purity to $\geq 98\%$ while reducing the efficiency 
to 93\%. Two additional cuts are imposed to 
reduce outside background by requiring 
that there be no tracks upstream 
of the first photon conversion (\gf) and that 
there be no hits associated with the 
tracks composing the \gf\  in the most upstream DC. 
The second photon conversion, \gs, 
occurs downstream. The two reconstructed 
photon momentum vectors enable one to 
determine the $\nu$-interaction vertex by 
extrapolating the vectors upstream and finding 
the coordinates of their distance of closest approach (DCA). 
The procedure defines the DCA-vertex with 
coordinates denoted as DCA-X, DCA-Y, and DCA-Z. 
The DCA-vertex resolution is well understood 
using ordinary $\nu$-interactions where the 
primary charged tracks composing the 
event vertex are ignored 
and the rest of the 
event is subjected to the \gf\ and \gs\ reconstruction.
The DCA-X and DCA-Y resolution is 
$\simeq 2.5$~cm. However, the DCA-Z resolution is poor, 
$\simeq 13$~cm.
This is expected since photons from a \cohp\ decay  
have a small opening angle, consequently  
their intersection in the Z-direction will be poorly determined. 
Finally, the angular resolution of the  \gf\ 
and \gs\ vectors is precise ($\simeq 5$~mrad) 
but the momentum resolution, 
as determined via  the curvature of the 
$e^-$ and $e^+$ tracks,  is poorer ($\simeq 13\%$) due to the 
bremsstrahlung losses. 
Therefore we have principally relied  
upon angular variables to determine the signal.
Table~\ref{tab-sel} summarizes the selection of events in 
the MC samples. 
The reconstruction efficiency of the \cohp\ signal 
is 7.8\% (the BK model yields 7.7\%.) 
Table~\ref{tab-sel} also shows that the NC-Resonance 
production contributes less than 1\% to the sample. In the following 
the resonance contribution is simply added to the NC-DIS 
component. 
The preselected data are subjected to identical cuts. Having 
identified the two photons, and having imposed the DCA-X/Y 
cuts, data can be compared with the respective predictions 
as shown in the Table~\ref{tab-finalsel}. 
Note that the fraction of events failing the DCA-Z 
cut is  larger in  data than those in  the \cohp\ and NC-DIS 
simulations.  
This is due to neutrinos interacting in material just 
outside the fiducial volume cut such as the
magnet, coil, etc.,  which are not simulated 
in the MC. Some of these 
interactions will also produce events 
with DCA-Z $\geq Z_{Min}$. The measurement of this  
background and the calibration of the NCDIS and \cohp\  
predictions are presented in the following section. 

\begin{table}\centering
{\small{

\begin{tabular}{||c||c||c|c||}
\hline
Cut                                &  \cohp-RS  &  NC-DIS  &  NC-Res  \\ 
\hline \hline      

Raw                                  &  1435.4      &   4743.2   &  1132.8    \\
No $\mu$-ID                   &  1435.4 &   4687.9   &  1125.7     \\ \hline      

\gf\ Fid-Cuts                        &  1373.0    &   4682.3   &  1030.4 \\

\gf\ $M_{ee}\leq50$~MeV   &   917.5    &   3664.9   &    27.2   \\

No Upstream Track         &   862.2    &   1717.7   &    23.8       \\
No Veto                             &   858.4     &   1659.5   &    23.7     \\ \hline      

\gs\ Fid-Cuts                        &   128.9   &    311.7   &     1.2    \\ 

\gs\ $M_{ee}\leq50$~MeV        &   117.5   &    236.7   &  1.1  \\  \hline      

$E_{\pi^0}\geq 0.5$~GeV           &   117.5    &    236.7  &   1.1   \\

DCA-$|X,(Y-5)|\leq 130$~cm      &   115.9   &    225.2   & 1.0   \\ \hline      

DCA-$Z \geq Z_{Min}$              &   112.6      &    222.5  &  1.0  \\ \hline      

DCA-$Z \leq Z_{Min}$                &     3.3         &      2.7  &    0.0  \\ 
\hline \hline

\end{tabular}
\caption{Selection of Exclusive 2-$\gamma$ Events in the MC Samples:  
The MC samples have been normalized as presented in Section~\ref{sec-sel}.} 
\label{tab-sel}
}}
\end{table}

\section{Extraction of the  \cohp\ Signal}
\label{sec-signal}

The extraction of \cohp\ signal is data driven. 
Monte Carlo simulations can neither reliably provide 
the normalization of the outside-background nor the 
normalization of the NC-DIS induced \piz\ where 
nothing else is visible 
nor the shape of the $\zeta$ variables.  
Distinct control samples in the data  
provide a measure of these backgrounds, including 
the integral and the shape of the variables 
relevant to this analysis. 

First we present the measurement of  
background induced by $\nu$-interactions outside 
the fiducial volume (OBG).
As shown in Table~\ref{tab-sel}, 
the fraction of MC events
in the fiducial region but with DCA-Z $\leq Z_{Min}$ 
is negligible.  The 169 data events that fail 
the DCA-Z cut (see Table~\ref{tab-finalsel}) are dominated by 
interactions upstream of the detector ($Z \leq Z_{Min}$); 
the contribution from the events entering from the sides 
give a small contribution 
($\leq 2\%$ of the background). This is for two reasons: 
first, since the transverse resolution of DCA-vertex is accurate 
to $\simeq \pm 3$~cm, the DCA-X and DCA-Y cuts largely eliminate  
these events; second, among the events relevant to 
the \cohp\ selection the two photons travel along the beam 
while particles entering the detector from the sides 
have much larger angles. 

\begin{figure}
\begin{center}
\includegraphics[width=0.9\textwidth]
{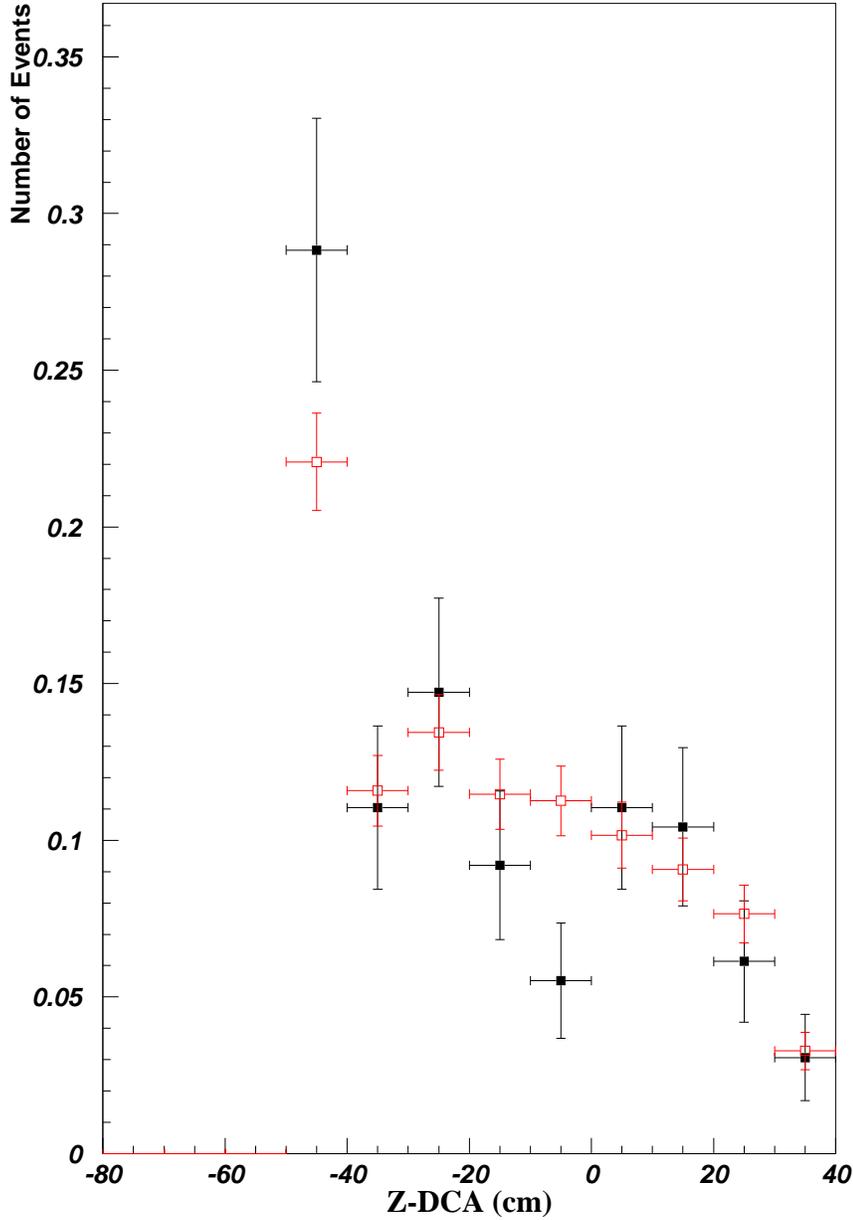}
\caption
{Comparison of the $Z$-DCA Distributions Failing DCA-Cut:
Shown are $Z$-DCA distributions of the \cohp\  
sample 
(solid-black) 
and that of  events originating from 
interactions upstream (open-red).
}
\label{fig-dca-comp-obg}
\end{center}
\end{figure}

The 169 events  failing  the DCA-Z cut (Table~\ref{tab-finalsel})  
are the key to providing the normalization for the 
outside-background (OBG).
To determine the OBG a different data sample
is selected in which a
vertex is reconstructed upstream of the detector 
($Z \leq Z_{Min}$). In this
control sample the primary tracks are then ignored 
and the events are subjected to the \cohp\ analysis.  
A total of 1378 events survive this selection of which 
451 (927) events have the DCA vertex  
within (outside) the fiducial volume.
Figure~\ref{fig-dca-comp-obg} compares the shape of the 
Z-distribution of the DCA of the 169 events that fail the DCA cut in
the \cohp\ signal sample with the 927 events that fail this cut in the 
control sample. The shapes agree well. 

We thus measure the normalized OBG prediction to be: 
$ \left [ 451/927 \right ] \times 169 = 82.2 \pm 6.9$ events.
The distributions of the OBG variables  (vertex position, \zt, \mgg, etc.)  
are measured using the two-photon data with 
DCA-Z$\leq Z_{Min}$ normalized to 82.2 events. 
Table~\ref{tab-finalsel} presents the calibrated OBG background.

Second, we present the measurement of the 
NC-DIS background. 
The NC-DIS component of the 2-\gam\  sample is 
selected using the kinematic variables. We 
use events with $M_{\pi^0} \geq 0.2$~GeV/$c^2$ or 
$\zeta_{\gamma 1 / \gamma 2} \geq 0.05$,
where the \cohp\ contribution 
is minimal, to obtain the normalization of the 
NC-DIS, 0.86,  with a 7.5\% statistical precision. 
The distributions of the NC-DIS variables  
predicted by the MC are corrected 
using the Data-Simulator (DS) technique:
first, NC events with a reconstructed primary 
vertex are selected from both data and MC; then, 
after removing  the primary tracks,  these events are subjected 
to the \cohp\ analysis; finally, the ratio  Data/MC 
provides the DS-correction. This correction is found to 
be unity within $\pm 10\%$.  
Table~\ref{tab-finalsel} presents the calibrated NC-DIS background.

\begin{table}\centering
{\small{

\begin{tabular}{|||c||c|c|c|c||c|||}
\hline
Cut                                &  \cohp-RS  &  NC-DIS  &  OBG   & Total  &  Data \\ 
\hline \hline      
DCA-$|X,(Y-5)|\leq 130$~cm      &   114.2   &  193.7   & 241.9 &  549.8 & 550  \\ \hline      
DCA-$Z \geq Z_{Min}$                &   110.9     &  191.4 &   82.2 & 384.5 & 381 \\ \hline      
DCA-$Z \leq Z_{Min}$                 &     3.3         &     2.3 &  159.7 & 165.3 & 169 \\ \hline \hline      

\end{tabular}
\caption{DCA-Cuts and the 2-$\gamma$ Samples: 
Data and predictions passing the DCA cuts are shown.  
The final calibration of the 
\cohp\ and background predictions are given in 
Section~\ref{sec-signal}.
} 
\label{tab-finalsel}
}}
\end{table}

Finally, we present the extraction of the \cohp\ signal which 
is based upon three variables:  \ztf, \zts, and \thfs, 
where \thfs\ is the opening angle between  \gf\ and \gs.  
The choice of variables is dictated by the resolution. 
The variables \ztf\ and \zts\ are correlated while 
\thfs\ displays no correlation with the former variables. 
A $\chi^2$ between data and prediction is defined
using two distributions: the two-dimensional \ztf\ and 
\zts\  distribution, and the \thfs\  distribution. The $\chi^2$ 
between the data and the prediction is minimized with respect to 
the \cohp\ normalization factor, $\alpha$. 
The expected numbers of OBG and NC-DIS events are determined 
as described above, and are kept fixed, while the simulated 
\cohp\ sample is normalized to 5000 generated events. The 
$\chi^2$ is minimized with respect to $\alpha$ which is varied 
between 0 and 2 in steps of 0.01. The minimum 
$\chi^2$, 45.1 for 44 degrees of freedom (DoF), is obtained for 
$\alpha = 0.985 \pm 0.113$. The probability of this fit 
is 0.44. 
Using the number of \cohp\ signal (112.6) 
in Table~\ref{tab-sel} and $\alpha = 0.985$, 
we extract the observed signal: $110.9 \pm 12.5$. 
The error is statistical and corresponds to a $\chi^2$ 
change by one unit.

To check if the two photon data can be explained using only 
OBG and NC-DIS component, we set the \cohp\ contribution 
to zero and fit for the normalization of OBG and NC-DIS --- 
their respective distributions being fixed by the data. The 
best $\chi^2$ was 80.3 for 43 DoF but neither 
the normalization nor any of the data distributions --- 
the \gf\ and \gs\ vertex positions, 
the DCA-vertex position, energy, $P_T$, \zt, \mgg, etc. --- 
are well described by this hypothesis. The probability of 
this fit is  $\leq$0.001.

\begin{figure}
\begin{center}
\includegraphics[width=0.9\textwidth]
{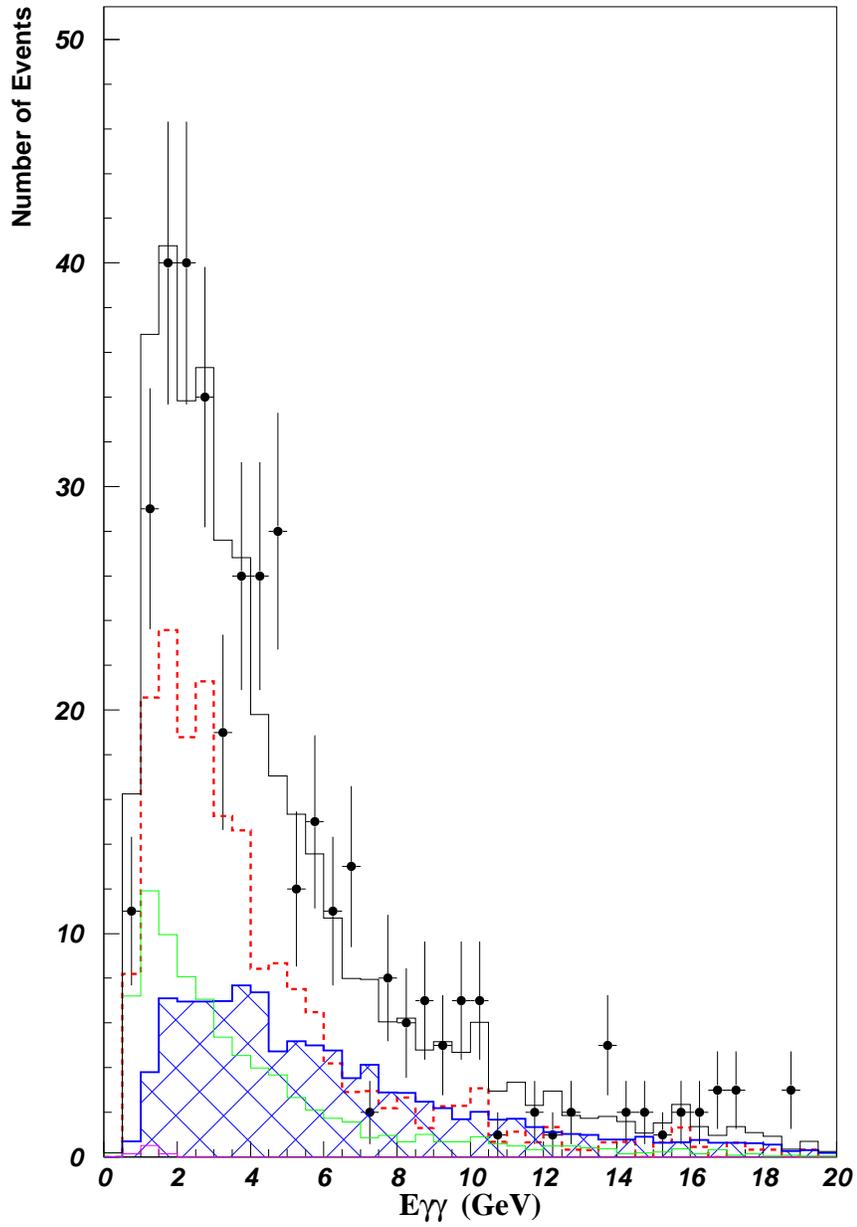}
\caption
{Comparison of the $E_{\gamma \gamma}$, defined as 
$E_{\gamma 1}+E_{\gamma 2}$, between data (symbol) and 
MC (\cohp\ in hatched blue, OGB in dot-dash green, NCDIS in 
dotted red, total in solid histograms).}
\label{fig-epi}
\end{center}
\end{figure}

\begin{figure}
\begin{center}
\includegraphics[width=0.9\textwidth]
{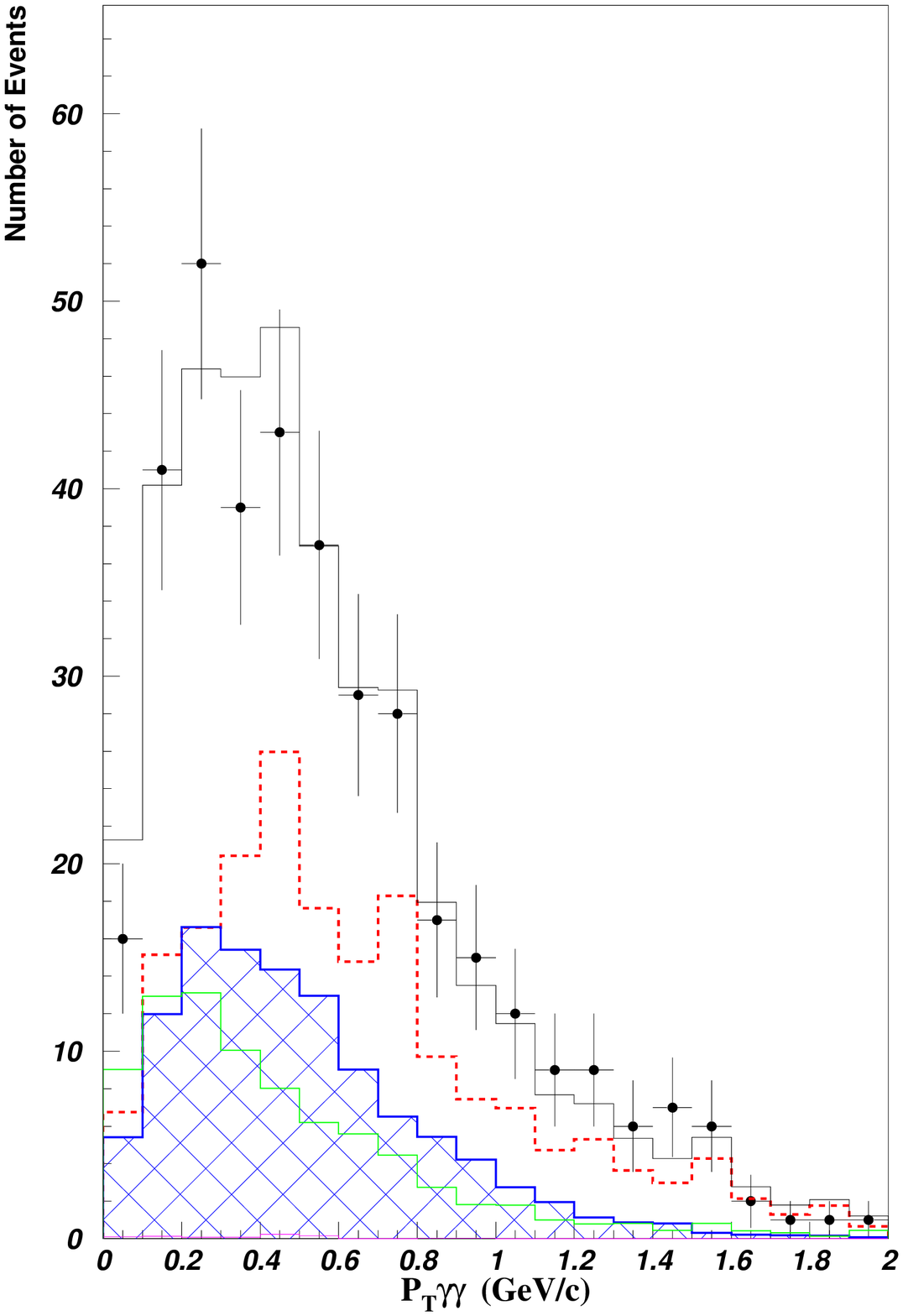}
\caption
{Data and MC Comparison of the $P_{T \gamma \gamma}$ Distribution.}
\label{fig-ptpi}
\end{center}
\end{figure}

\begin{figure}
\begin{center}
\includegraphics[width=0.9\textwidth]
{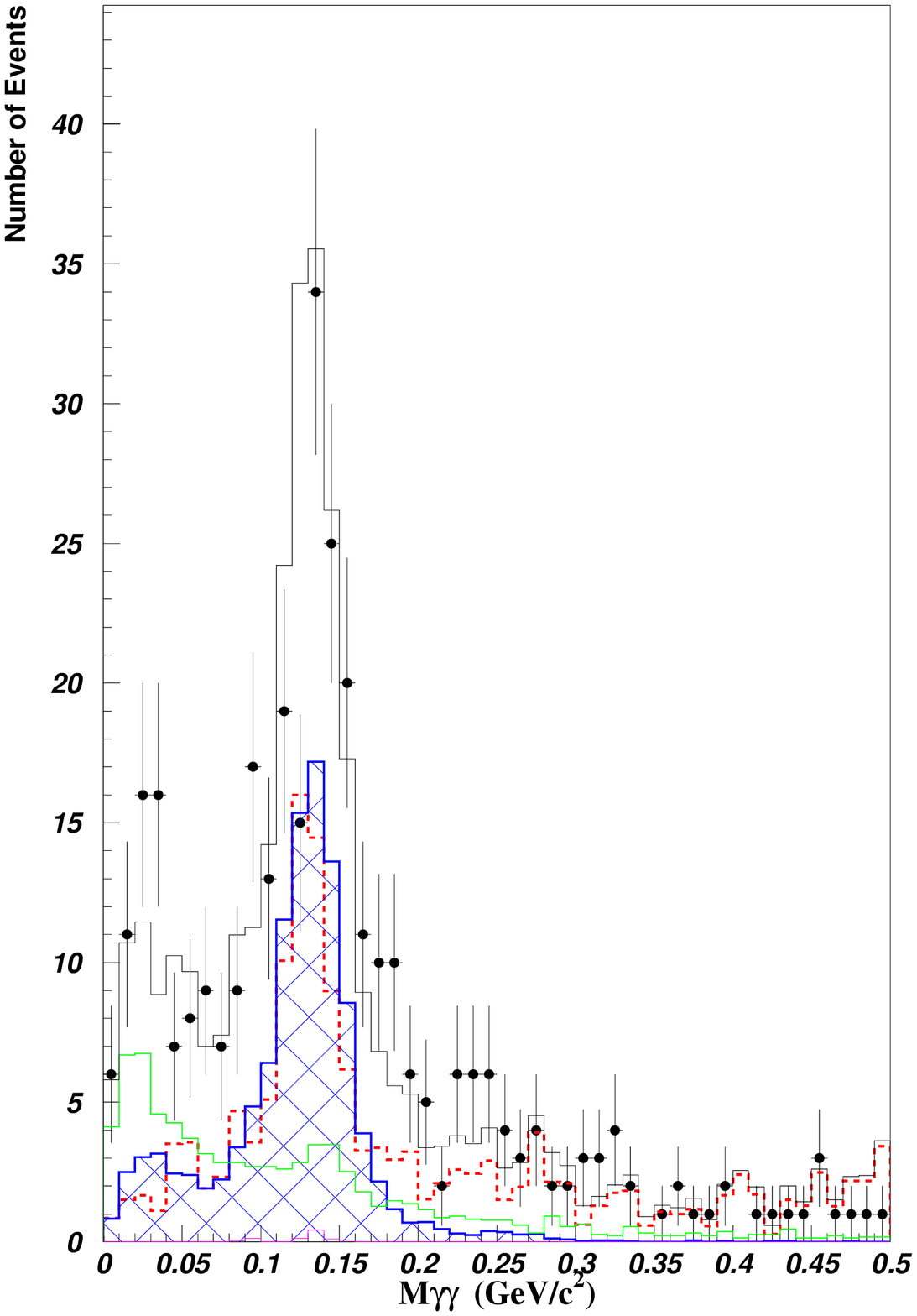}
\caption{Data and MC Comparison of the $M_{\gamma \gamma}$ Distribution.}
\label{fig-mpi0}
\end{center}
\end{figure}

\begin{figure}
\begin{center}
\includegraphics[width=0.9\textwidth]
{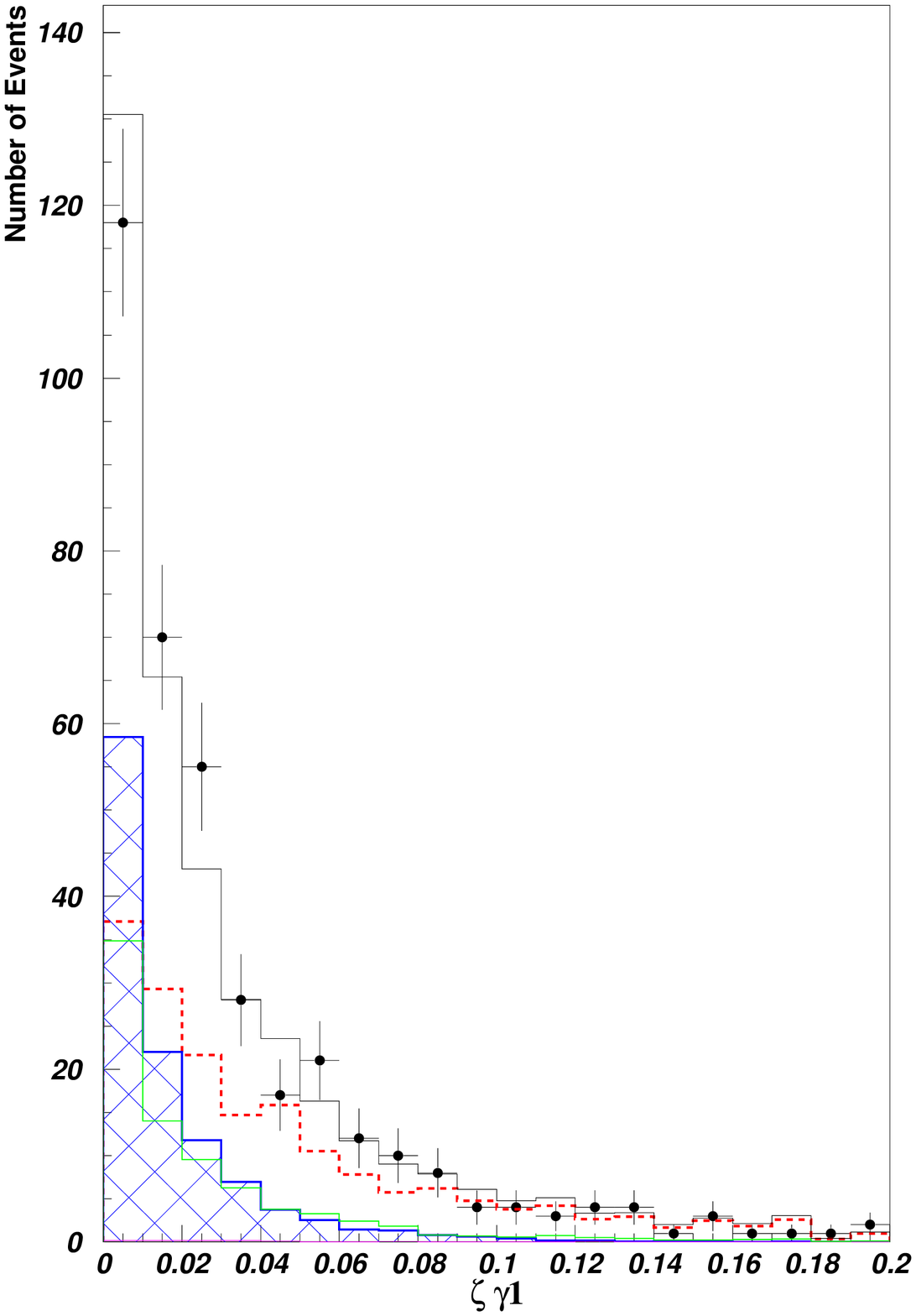}
\caption{Data and MC Comparison of the \ztf\ Distribution.}
\label{fig-zeta1}
\end{center}
\end{figure}

\begin{figure}
\begin{center}
\includegraphics[width=0.9\textwidth]
{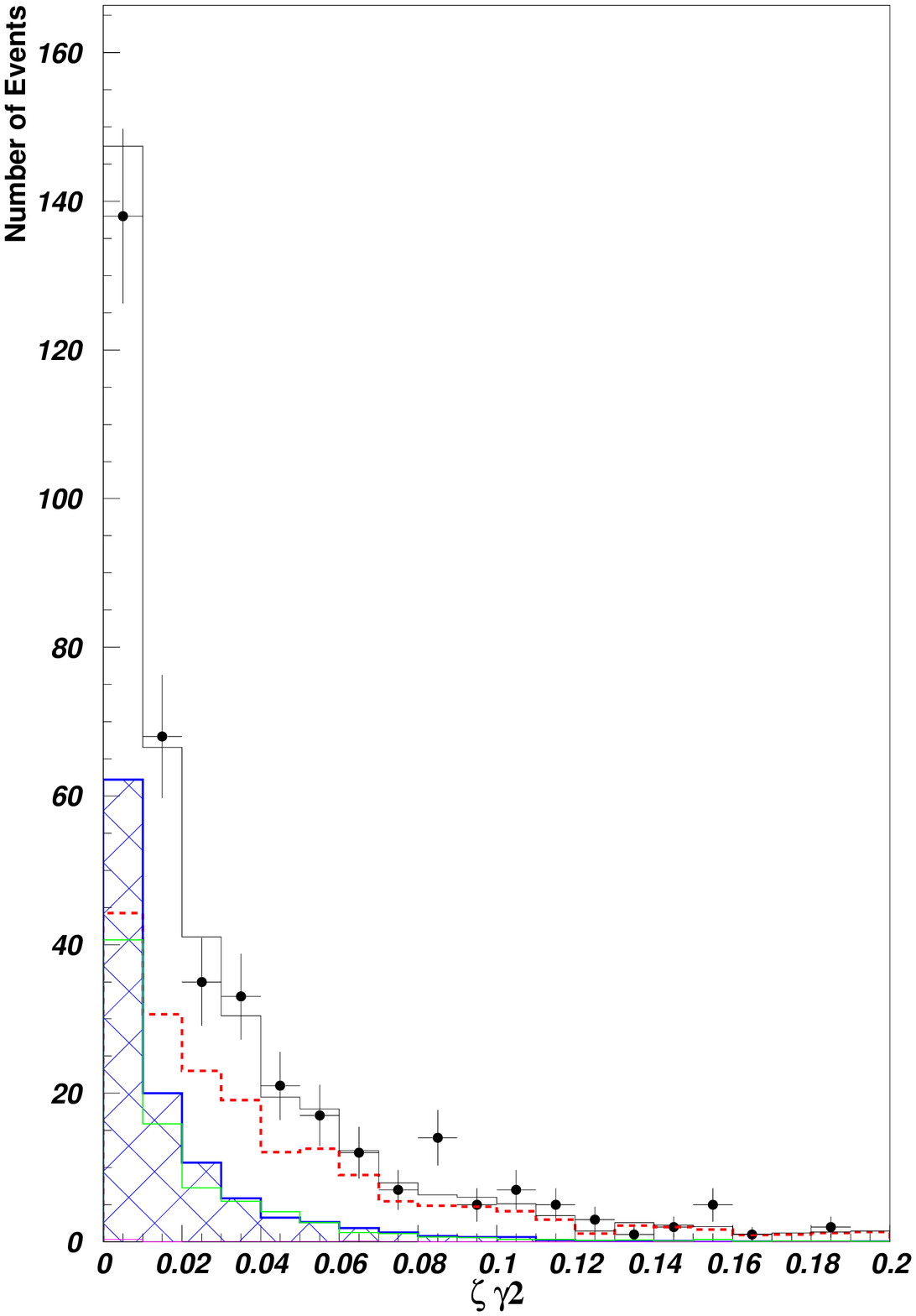}
\caption{Data and MC Comparison of the \zts\ Distribution.}
\label{fig-zeta2}
\end{center}
\end{figure}

\begin{figure}
\begin{center}
\includegraphics[width=0.9\textwidth]
{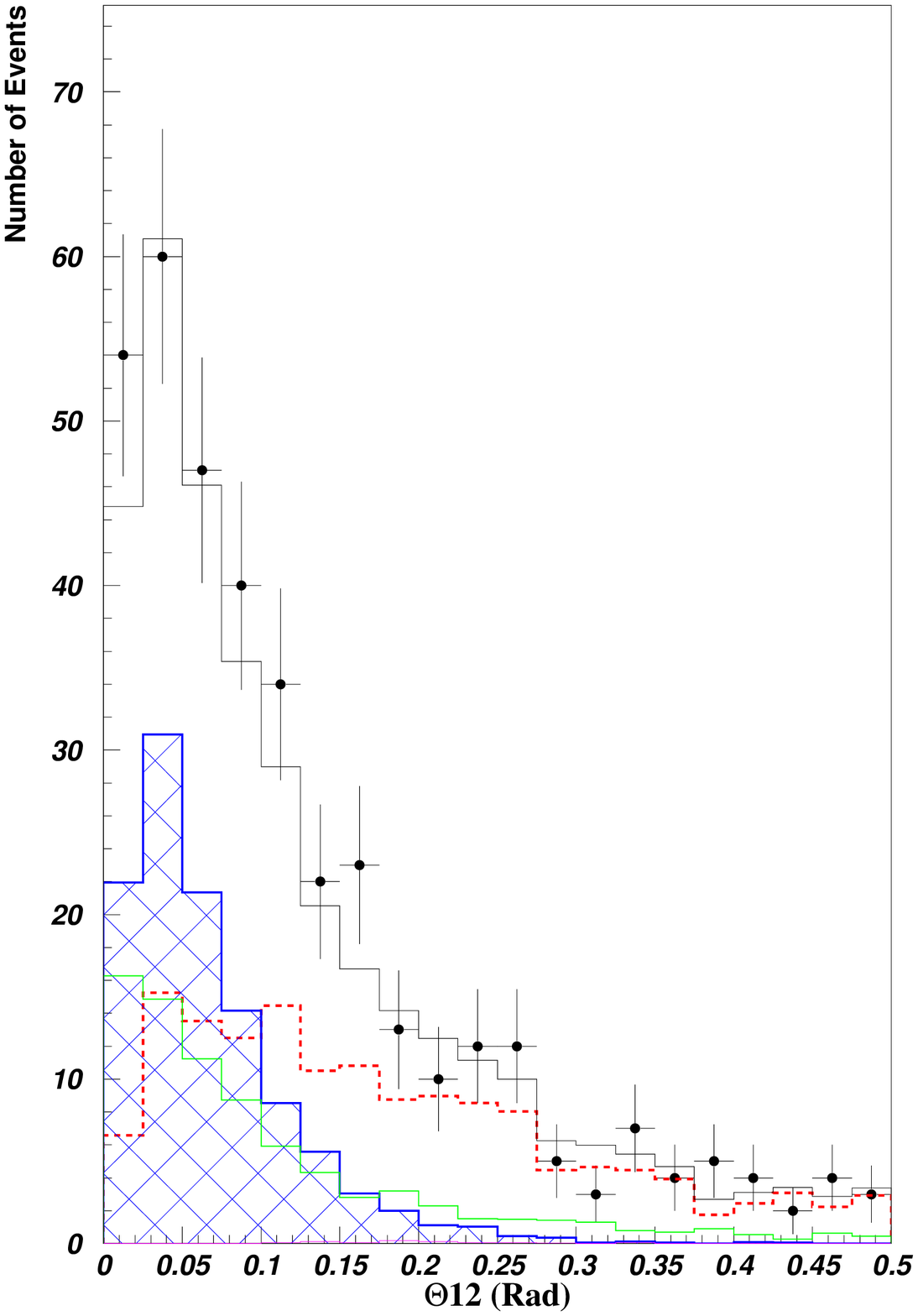}
\caption{Data and MC Comparison of the \thfs\ Distribution.}
\label{fig-theta12}
\end{center}
\end{figure}

\begin{figure}
\begin{center}
\includegraphics[width=0.9\textwidth]
{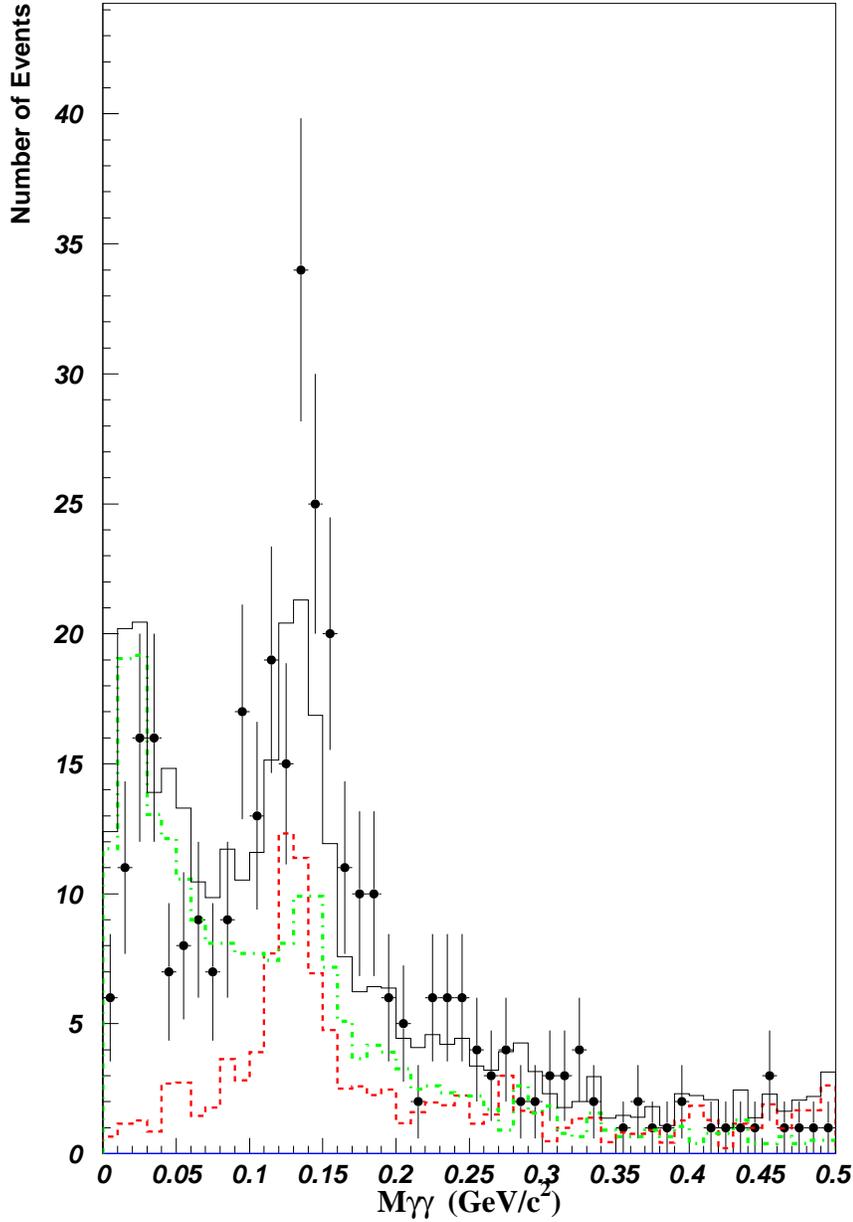}
\caption{Comparison of the $M_{\gamma \gamma}$ Distribution between 
data and the best fitted (OBG+NC-DIS) with \cohp\ set to zero. }
\label{fig-mgg-bkg}
\end{center}
\end{figure}

Having determined all the components of the 
2-\gam\ sample, Table~\ref{tab-finalsel} compares 
the final predictions with the data. Below 
we present a comparison of a set of salient variables between 
data in symbols and expectation --- 
DS-corrected NC-DIS in red-dotted histogram, 
OBG in green-histogram, the 
\cohp\ signal in blue-coarsely-hatched histogram, 
and the total expectation (MC) in black histogram. 
Figure~\ref{fig-epi} and Figure~\ref{fig-ptpi} 
compare the 
$E_{\gamma \gamma}$, defined as $E_{\gamma 1}+E_{\gamma 2}$, 
and $P_{T{\gamma \gamma}}$ distributions. 
Figure~\ref{fig-mpi0} compares the 
invariant mass distribution computed using the \gf\ and \gs\ vectors. 
Figure~\ref{fig-zeta1}  and Figure~\ref{fig-zeta2} 
compare the \ztf\ and \zts\ distributions; 
and Figure~\ref{fig-theta12} compares the \thfs\ distribution. 
The agreement between data and MC for the variables 
is satisfactory. 
For illustration, in Figure~\ref{fig-mgg-bkg} 
we present the comparison 
of the \mgg\ distribution between data  and the best 
fitted (OBG+NC-DIS) prediction with \cohp\ set to zero: 
here the Data-vs-MC $\chi^2$ increases by 12 units compared to 
the Figure~\ref{fig-mpi0}. 

\clearpage \newpage

\clearpage \newpage

\clearpage \newpage

\clearpage \newpage

\clearpage \newpage

\clearpage \newpage

\section{Systematic Uncertainties }
\label{sec-syst}

The principal source of systematic error in the measurement 
of the \cohp\ cross section comes from 
the error in determining the NC-DIS induced 
contribution to the 2-\gam\ sample. 
The 7.5\% error in the NC-DIS contribution 
translates to 7.0\% in the signal. 
Since the OBG is entirely determined by the 
169 events that fail the DCA-cut, its contribution 
to the \cohp\ signal is computed to be 5.4\%. 
The error 
in the \piz\ reconstruction efficiency is estimated 
to be 2.7\% determined using $\gamma$-conversions 
from standard DIS interactions. 
Finally, the error in the absolute flux determination 
is determined to be 2.5\% which comes about as 
follows: the error is 2.1\% for $E_\nu \geq 30$ GeV, 
2.6\% for $10 \leq E_\nu \leq 30$~GeV, and 
4.0\% for $2.5 \leq E_\nu \leq 10$~GeV as 
determined in  ~\cite{NOMAD-XSEC}; these errors 
are folded in with the \cohp\ cross-section as a 
function of $E_\nu$ yielding an overall flux normalization 
error  of 2.5\%. 
These errors are summarized in Table~\ref{tab-errors}. 

\begin{table}
\begin{center}
 \begin{tabular}{|||c||c|||} \hline
Source                                 & Error  \\ \hline
NC-DIS                                & 7.0\% \\
OBG                                     & 5.4\% \\
\piz\ Reconstruction              & 2.7\% \\
Absolute Normalization        & 2.5\% \\ \hline
Total                                      & 9.5\% \\ \hline
\end{tabular}
\caption{Systematic Uncertainties in the \cohp\ Cross Section.}
\label{tab-errors}
\end{center}
\end{table}

\section{Result}
\label{sec-final}
Using the RS model,  the \cohp\ reconstruction efficiency is 
estimated to be 2.27\%. This value is the product of the 
fraction of \cohp\ events that trigger the apparatus (29.0\%), 
and the reconstruction efficiency (7.8\%). 
The $\nu$-sample is dominated by the \nm-interactions. 
The \cohp\ sample is corrected for the small 
contribution from other neutrino species to 
yield a pure \nm-contribution. 
The correction factor to account for the 
\anm, \ne, and \ane\ contributions to the \cohp\ 
interactions is 0.94. The factor takes into account 
the different energy spectra for the different $\nu$-flavors 
(we assume that the $\nu$ and 
$\bar \nu$ induced \cohp\ cross sections are the 
same).  The error in the \cohp\ cross section 
due to this 6\% correction is $\leq 0.6\%$ and 
is deemed negligible in this analysis. 
Thus the \nm-induced \cohp\ events are  
{\boldmath $4630 \pm 522 (stat) \pm 426 (syst)$}  events. 
The number of fully corrected \nm-CC in the same fiducial volume 
is measured to be $1.44 \times 10^{6}$. Our result is:

\begin{equation} 
\frac {\sigma (\nu {\cal A} \rightarrow \nu {\cal A} \pi^0)}
{\sigma (\nu_\mu {\cal A} \rightarrow \mu^- X)} = 
\left [ 3.21 \pm 0.36(stat) \pm 0.29(syst)  \right ] \times 10^{-3}
\label{eq-ccrat}
\end{equation}

Using the measured inclusive \nm-CC cross-section 
from ~\cite{NOMAD-XSEC} as a function of $E_\nu$, 
the absolute cross section of \cohp\ production 
for ${\cal A}=12.8$ 
at the average energy of the neutrino flux $E_\nu = 24.8$~GeV 
is determined to be:

\begin{equation} 
\sigma (\nu {\cal A} \rightarrow \nu {\cal A} \pi^0) = 
\left [ 72.6 \pm 8.1(stat) \pm 6.9(syst)  \right ] \times 10^{-40} 
cm^2/{nucleus} 
\label{eq-sigcohp}
\end{equation}

The measurement agrees with the RS prediction of 
$\simeq (78 \times 10^{-40}) cm^2/nucleus $ using 
${\cal A}=12.8$ and the CERN-SPS flux. 
A comparison of the NOMAD measurement of 
the \cohp\ with other published measurements is summarized 
in Table~\ref{tab-cohp-expt-sum}. 

To summarize, we have presented an analysis of 
the \cohp\ interaction in the \nm-NC using the 
two reconstructed photons in the final state.  
This is the most precise measurement of the \cohp\ process. 

\begin{table}
 \begin{tabular}{|||c||c|c||c|c|||} \hline
Experiment                                 & ${\cal N}$ucleus  & Avg-$E_\nu$ & $\sigma (Coh \pi^0)$ & \cohp/\nm-CC  \\ 
                                                   &                  &  GeV       & $10^{-40} cm^2/{\cal N}ucleus$ & $10^{-3}$ \\ \hline
Aachen-Padova ~\cite{EXAP}    & 27            &  2             & $(29 \pm 10)$ &  \\
Gargamelle ~\cite{EXGGM}       & 30            &  2             & $(31 \pm 20)$ & \\  
CHARM ~\cite{EXCHARM}        & 20            & 30            & $(96 \pm 42)$ & \\
SKAT ~\cite{EXSKAT}                & 30            &   7            & $(79 \pm 28)$ & 
$(4.3 \pm 1.5)$ \\
15' BC ~\cite{EX15FT}               & 20            & 20            & & $(0.20\pm0.04)$ \\
NOMAD                                      & 12.8         & 24.8            & $(72.6\pm 10.6)$ & 
$(3.21\pm 0.46)$ \\ \hline \hline 
\end{tabular}
\caption{Compilation of \cohp\ Measurements: 
We point out that Ref.~\cite{Kopeliovich:1992ym}  cites a value of 
$(2.0 \pm 0.4) \times 10^{-3}$ for \cohp/\nm-CC as attributed to 
~\cite{EX15FT}.
}
\label{tab-cohp-expt-sum}
\end{table}

\section*{Acknowledgments}
We gratefully acknowledge the CERN SPS staff for the magnificent 
performance of the neutrino beam. The experiment was supported 
by the following agencies: 
ARC and DIISR of Australia; IN2P3 and CEA of France, BMBF of 
Germany, INFN of Italy, JINR and INR of Russia, FNSRS of 
Switzerland, DOE, NSF, Sloan, and Cottrell Foundations of 
USA, and VP Research Office of the University of South Carolina.

\end{document}